# A Parallel Encryption Algorithm for Block Ciphers Based on Reversible Programmable Cellular Automata

Debasis Das and Abhishek Ray

**Abstract**—A Cellular Automata (CA) is a computing model of complex System using simple rule. In CA the problem space into number of cell and each cell can be one or several final state. Cells are affected by neighbours' to the simple rule. Cellular Automata are highly parallel and discrete dynamical systems, whose behaviour is completely specified in terms of a local relation. This paper deals with the Cellular Automata (CA) in cryptography for a class of Block Ciphers through a new block encryption algorithm based on Reversible Programmable Cellular Automata Theory. The proposed algorithm belongs to the class of symmetric key systems.

**Index Terms**—Cellular Automata (CA), Programmable Cellular Automata (PCA), Reversible Programmable Cellular Automata (RPCA), Cryptography, Block Ciphers.

## 1 Introduction

The popularity of Cellular Automata [2] can be traced to their simplicity and are used for the modelling of complex system using simple rule [1]. This can be viewed as a simple model of a spatially extended decentralized system made up of a number of individual components, called as cell. The communication between cells is performed using simple rules. Each individual cell is in a specific state, which changes over time depending on the state of its local neighbours. The overall structure can be viewed as parallel processing device. The state of each cell is updated simultaneously at discrete time based on the states in the neighbourhood at the preceding time step.

A CA is said to be reversible CA in the sense that the CA will always return to its initial state. In CA each cell is affected by its left and right neighbourhood. A cell composition is formed by the cell itself and its left and right neighbours and is represented through a parameter r called as radius. If the radius is 1 then the composition 'm' include (2*r) + 1 cells. This makes $n = 2^m$ possible pattern and $2^n$ possible rules i.e. if r=1, m = 2*1 + 1 = 3, $n = 2^3 = 8$ possible pattern, $2^8$ means 256 rules.

The finite CA is used with cyclic boundary conditions. Such CA can be treated as ring. When the same rule is applied to all cells, the CA is said to be a Uniform CA, otherwise it is called non-uniform.

The block encryption algorithm used to compute the next cell state based on CA local rules. The local rules are classified based on existing categories [3], [5], [6]. A CA is called Programmable CA if it employs some control signal.

The main aspects of information security [4] due to rapid Information Technology application in various fields are privacy, data integrity, authentication, and non-repudiation. Encryption and decryption are two complementary operations satisfying the demand of privacy. Cryptographic techniques are divided into two categories [4] symmetric key and public key. There are two classes of symmetric key encryption schemes [4] block ciphers and stream ciphers. This paper deals with symmetric key block ciphers and considers both issues of encryption and decryption. CA has been used so far in both symmetric key and public key cryptography. CA based public cipher was proposed by Guan [7]. Security of block encryption algorithm was based on difficulty of solving a system of nonlinear polynomial equations. Stream CA based encryption algorithm was first proposed by Wolfram [8]. The idea was to use CA as a pseudo-random number generator (PRNG). The generated sequence was combined using XOR operation with the plain text. The result of that operation formed the cipher text. The secret key was the initial state of CA. In the decryption process some pseudo-random sequence needed to be regenerated using the secret key and then combined with the cipher text.

• *F.A. Debasis Das, School of Computer Engineering KIIT University.*
• *S.B. Abhishek Ray, School of Computer Engineering KIIT University.*

## 2 Cellular Automata

### 2.1 Simple Cellular Automata

Cellular automata are a collection of cells that each adapts one of a finite number of states. Single cells change in states by following a local rule that depends on the environment of the cell. The environment of a cell is usually taken to be a small number of neighbouring cells. Figure 1 shows two typical neighbourhood options.

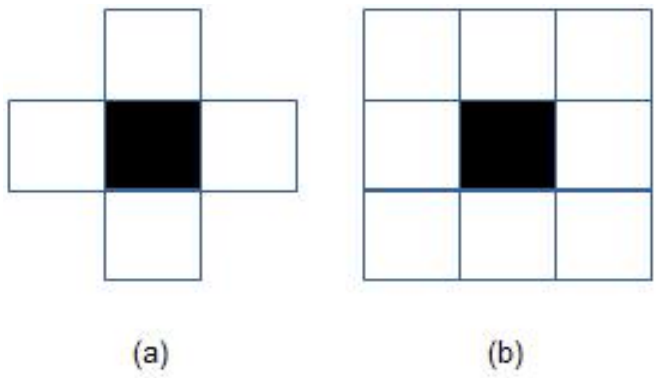

**Figure1 (a) Von Newman (b) Moore Adjacency**

One-dimensional CA is an array of cells. Each cell is assigned a value over some state alphabet. CA is defined by five parameters: size, initial state, neighbourhood, rule and boundary conditions. Size defines number of cells. All cells update its value synchronously in discrete time steps accordingly to some rule. Such rule is based on the state of the cell itself and its neighbourhood:

$$S_i(t+1)= R(S(i-r)(t),......,S(i-1)(t), S(i)(t),S(i+1)(t) ........S(i+r)(t))$$

where $S_{(i)}(t)$ is a value of $i^{th}$ cell (the state of a cell) in step $t$ and $r$ is a radius of the neighbourhood. When dealing with finite CA, cyclic boundary conditions are usually applied which means that CA can be treated as a ring. Changing values of all cells in step $t$ is called CA iteration. Before the first iteration can take place some initial values must be assigned to all cells. This is called the initial state of CA. By updating values in all cells, the initial state is transformed into a new configuration. When each cell updates its state according to the same rule, CA is said to be uniform. Otherwise it is called non-uniform CA. The total number of rules considered for different reversible CA classes are presented in [9]. This paper presents the idea of using reversible CA class presented by Wolfram [12]**.** In this class rule depends not on one but on two steps back:

$$S_i(t+1)= R(S(i-r)(t),...,S(i-1)(t), S(i)(t), S(i)(t-1)S(i+1)(t)....S(i+r)(t))$$

In the elementary CA value $S_{(i)}(t+1)$ of $i^{th}$ cell in configuration $t+1$ depends on the value of the state of itself and $r$ of its neighbours in configuration $t$. In this reversible class additional dependency is added: the value of the central cell $S(i)(t-1)$ in step (t-1) is considered. Such a rule can be simply constructed by taking elementary CA rule and adding dependency on two steps back.

Definition of the rule is now composed of two elementary rules. The first one is defining state transition in case when in step *t-1* cell was in a state 1, and the second one when the cell was in the state 0. If two elementary rules: rule 236 and rule 19. These two rules are complementary to each other. Knowing one value it is possible to calculate the second one using the following formula:

$R2= 2^d-R1-1$

Where $d=2^{(2*r+1)}$, and r is radius of the neighbourhood,R1 given rule and R2 complement rule. One interesting property of Reversible Cellular Automata is that the same reversible Rule is used in both in Forward and Backward iteration. Science a reversible rule dependent now on two steps back, an initial configuration must be composed of two successive configurations $q_0$ and $q_1$.

### 2.2 Programmable Cellular Automata

P. Anghelescu, S. Ionita and E. Sofron et al [12] have been explored Programmable Cellular Automata. A CA is called Programmable CA if it employs some control signals. By specifying values of control signal at run time programmable CA can implement various function dynamically.

The PCA was firstly introduced in [14], where the control logic of each cell not fixed but controlled by a number of control signals such that different function (i.e. Rule) can be realized on the same structure. As the matter of fact, PCA are essentially a modified CA structure.

The total number of rules for radius r neighbourhood is $2^n$ where $n=2^{(2*r+1)}$. In this paper one dimensional PCA defined over binary state alphabet (cells can be either in state 0 or 1) with neighbourhood size two and three is used.

### 2.3 Reversible Cellular Automata

By applying a rule to each cell $S_{(i)}(t)$ of the configuration $q_{(i)}(t)$ a new configuration $q_{(i)}(t+1)$ is obtained. This transformation can also be defined by a global transition function, which takes as an input configuration $q_{(i)}(t)$ and results in a successive configuration $q_{(i)}(t+1)$. A CA is reversible if and only if the global transition function is one-to-one that is if every configuration not only has one successor but also has one predecessor.

Reversible rules that could be useful in cryptography should meet the following criteria: they should be numerous and they should exhibit complex behaviour. When analyzing elementary CA it turns out that only a small number of rules have the property of being reversible. For example, among all 256 of radius 1 CA of only six are reversible. This is why class of CA with rules specially created to be reversible is.

## 3 Survey of Existing Encryption Algorithm

### 3.1 Encryption Algorithm Based Programmable CA

The block diagram of PCA encryption systems is presented in Figure. 2. The rules specify the evolution of the CA from the neighbourhood configuration to the next state and these are presented in Table-1.

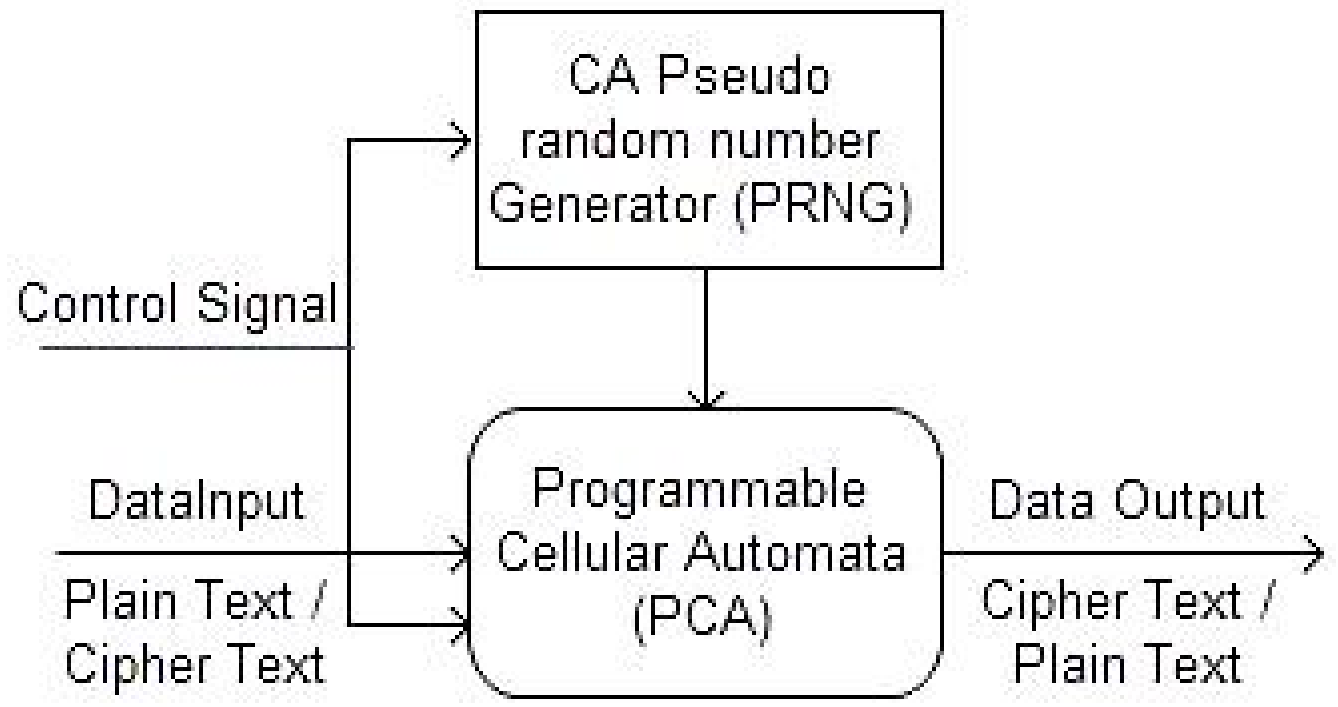

**Figure 2: Block Diagram of PCA Encryption System**

The corresponding combinational logic of rules 90 and 150 for CA can be expressed as follows:

$$Rule90 : a_i(t+\ ) = \imath_{(i-\ )}(t) \oplus {}_{(i)}(t) \oplus {}_{(i+\ )}(t) \quad (1$$

$$Rule150 : a_i(t+\ ) = \imath_{(i-\ )}(t) \oplus {}_{(i+\ )}(t) \quad (2$$

The simple CA is used to provide real time keys for the block cipher. The operation of the simple CA can be represented by the state transition graph.

**Table 1: The rules That Updated The next state of the CA cells**

| Rule | 111 | 110 | 101 | 100 | 011 | 010 | 001 | 000 |
|---|---|---|---|---|---|---|---|---|
| 90 | 0 | 1 | 0 | 1 | 1 | 0 | 1 | 0 |
| 150 | 1 | 0 | 0 | 1 | 0 | 1 | 1 | 0 |
| 153 | 1 | 0 | 0 | 1 | 1 | 0 | 0 | 1 |
| 102 | 0 | 1 | 1 | 0 | 0 | 1 | 1 | 0 |
| 195 | 1 | 1 | 0 | 0 | 0 | 0 | 1 | 1 |
| 51 | 0 | 0 | 1 | 1 | 0 | 0 | 1 | 1 |
| 60 | 0 | 0 | 1 | 1 | 1 | 1 | 0 | 0 |

Each node of the transition graph represents one of the possible states of the CA. The directed edges of the graph correspond to a single time step transition of the automata.

By considering a 4 bit hybrid CA under the null boundary condition the rule vector is <90, 150, 90, and 150>. The transition graph is shown in Fig-3.

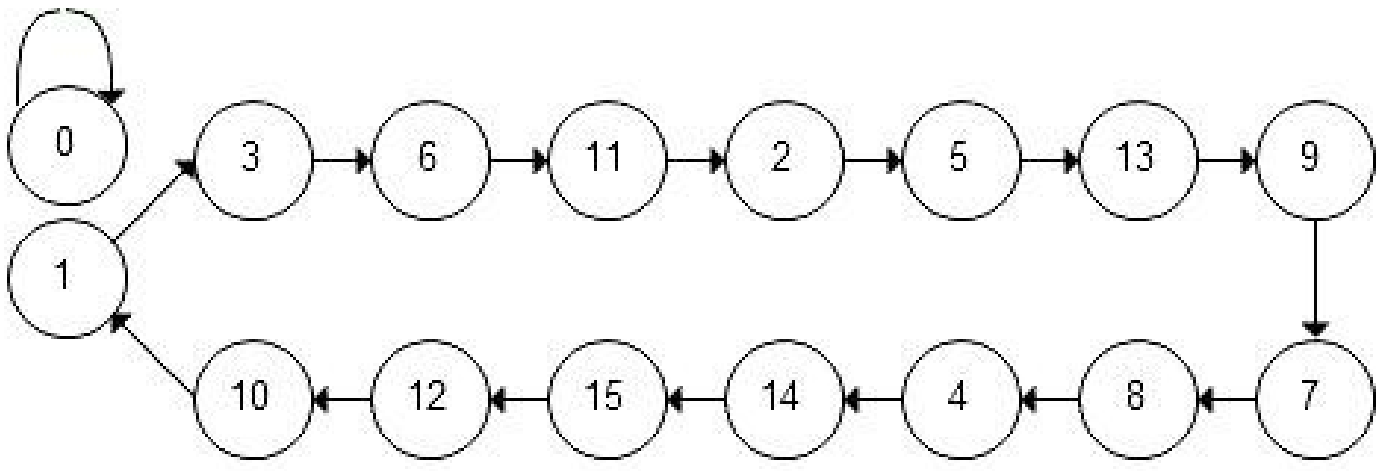

**Figure 3 The State Transition Diagram CA**

The combinational logic of rules 153, 102, 195, 51 and 60 [1] for the programmable CA can be expressed as follows:

$$Rule153 : a_i(t+1) = a_i(t) \odot a_{i+1}(t) \quad (3)$$

$$Rule102 : a_i(t+1) = a_i(t) \oplus a_{i+1}(t) \quad (4)$$

$$Rule195 : a_i(t+1) = a_{i-1}(t) \odot a_i(t) \quad (5)$$

$$Rule60 : a_i(t+1) = a_{i-1}(t) \oplus a_i(t) \quad (6)$$

$$Rule51 : a_i(t+1) = \neg a_i(t) \quad (7)$$

The rules 153, 195, 51, 60, 102, 204 are used in the block chipper [4] scheme.

**Algorithm 1: Null boundary State Transition for CA:**

Input : Rule Vector Size (n) and CA State Size
Output : State Transition Diagram for CA
Step 1 : Enter The Initial State of CA
Step 2 : Convert decimal value to binary and store in an Array
Step 3 : for j=1 to $2^n$
Step 4 : for i=1 to n
Step 5 : Apply the corresponding rule on the $i_{th}$ Cell
Step 6 : Store the next state value.
Step 7 : Convert binary to decimal value
Step 8 : End of loop2
Step 9 : End of loop1
Step 10 : Stop.

The encryption algorithm present in this paper is constructed using programmable CA based on rules 51, 195 and 153. The rules specify the evolution of the CA from the neighbourhood configuration to the next state and these are represented in Table 1.

Considering the CA with rule vector <153, 153, 153 and 153> under null boundary condition, the cycle is obtained as shown in the Fig-4.

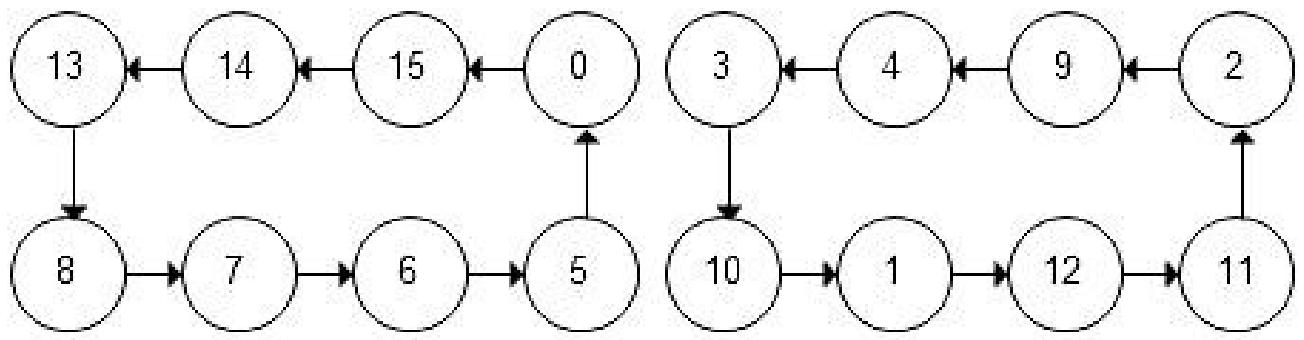

**Figure 4 State transition diagram of a CA (Using Rule Vector <153,153,153,153>)**

In Figure 4, the CA has two equal length cycles and each cycle has a cycle length 8, which is a basic requirement of the enciphering process.

### Algorithm2: Enciphering and Deciphering Process

Input : Given Plain Text/Cipher Text
Output : Cipher Text/Plain Text
Step 1 : Create State Transition Diagram of Cycle length using rule vector and apply the Corresponding rule.
Step 2 : Insert the value of plain text into original state of CA.
Step 3 : If it is goes to its immediate state after four cycles then
Step 4 : Plain Text is enciphered into cipher text.
Step 5 : else After running another four cycle the immediate state return back to its original state.
Step 6 : The cipher text is deciphered into plain text.

A circuit with two control signals C1 and C2 are shown in Fig-4. When C1 is logic 0 then rule 51 is applied to the cell. When C1 is logic 1 and C2 is logic 0 then the cell is configure with rule 195. When C1 and C2 are both logic 1 then the cell configure with rule 153.

Considering the CA with rule vector (51, 51, 195, 153) under null boundary conditions, we obtain the cycle as shown in the Figure 5.

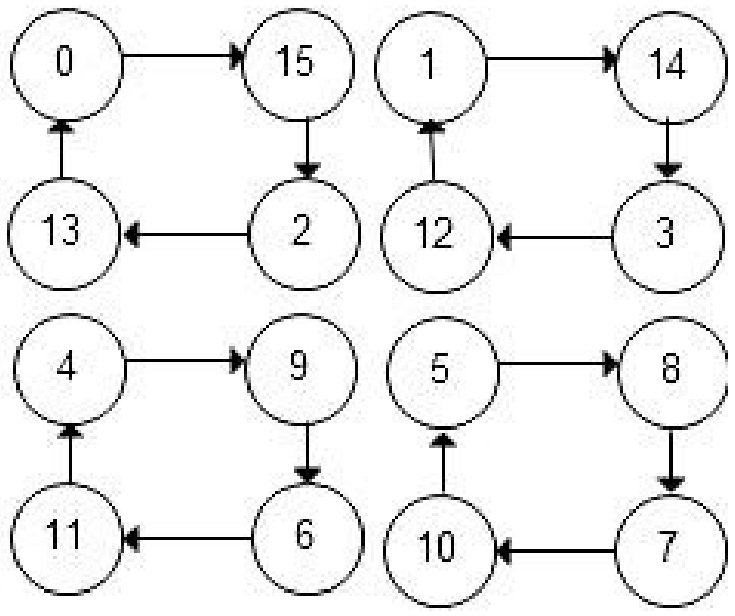

**Figure 5 State transition diagram of a CA (Using Rule Vector <51,51,195,153>)**

Cellular Automata used for modelling of complex system using simple rule. The

Selection of rule configuration reduces the circuit complexity. Block Encryption Algorithm are chosen PCA because reduce the circuit complexity. So, the diagram is not complex in Figure 6 and the corresponding truth table is simple in Table 2.

In Figure 6 the CA has four equal length cycles, each cycle has a cycle length 4. Considering this PCA as an enciphering function and defining a plaintext as its original state it goes to its immediate state after two cycles which is enciphering process. After running another two cycles, the immediate state returns back to its original state which deciphers cipher text into plain text ensuring deciphering process.

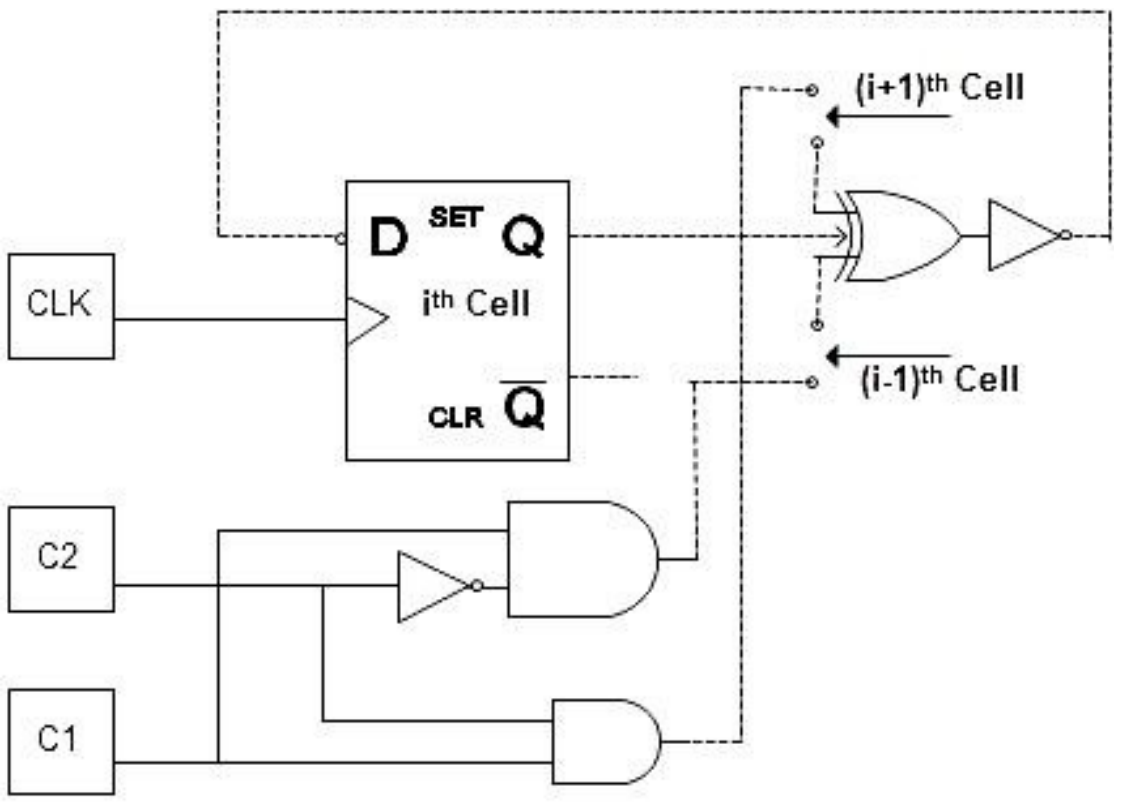

**Figure 6 PCA structure with rule 51,153,195**

**Table 2: Selection of Rules 51,195,153**

| C1 | C2 | Rules Applied |
|---|---|---|
| 0 | 0 | 51 |
| 0 | 1 | 51 |
| 1 | 0 | 195 |
| 1 | 1 | 153 |

## 4 Proposed Model Based on Reversible Programmable Cellular Automata:

### 4.1 Motivation

A CA is said to be reversible PCA in the sense that the CA will always return to its initial state using control logic. Reversible PCA means that not only forward but also reverse iteration is possible. Using a reversible rule it is always possible to return to an initial state of a CA at any point. If one rule is used for forward iteration then another rule that is reversible to the first one is used for the backward iteration. Using this type of rules for encryption process is show in the Figure 7.

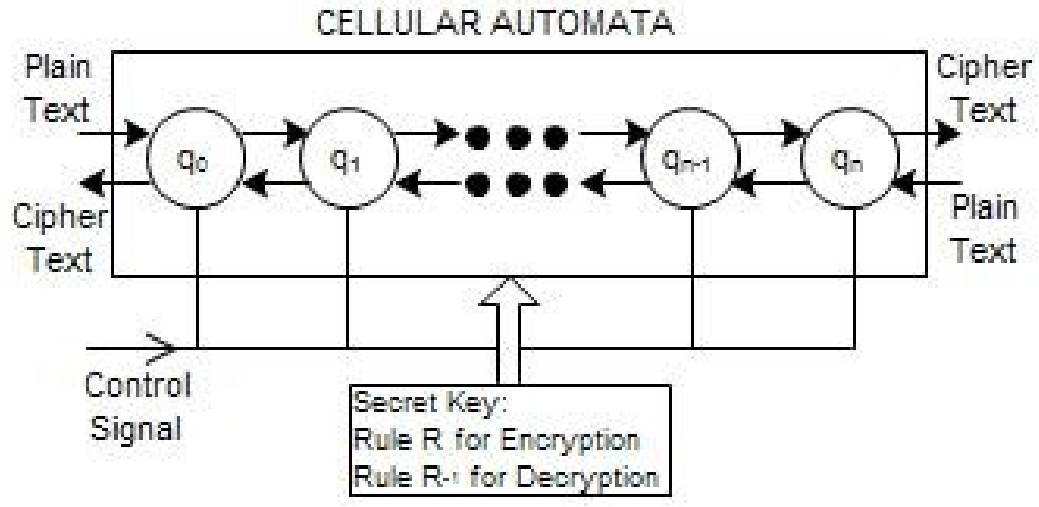

**Figure 7: Encryption/Decryption scheme using RPCA**

In Figure 6 the plain text will composed the initial state of the PCA means $q_0$ state of the above Reversible PCA and final state of PCA is $q_n$. Insert the value of plain text into $q_0$ state of PCA and received the output (cipher text) from state $q_n$. The Encryption process is then done by the forward iteration of the Reversible PCA. Decryption is based on the Reverse process, so cipher text is used as the initial state of the Reversible PCA and then Reversible PCA iterated backward process for the same number of steps than for encryption. The secret key is composed of two rules: the first rule is used for encryption and the second one is the rule reverse of the first one and is used for the decryption using the control signal.

Reversible PCA means the PCA will always return to its initial state. The Reversible CA rules satisfy the following criteria: a) rules are numerous and b) rules are exhibiting complex behaviour. By analyzing the elementary CA it turns out that only a small number of rules have the property being reversible. In radius 1 elementary CA out of 256 rules only six rules are reversible. The behaviour of the reversible rule is simple.

The reverse for the rule 15 is rule 240, rule 51 is rule 204 and rule 85 is rule 170. If two elementary CA rules are composing a reversible then one depend on each other. By knowing one of these rules other rules can be calculated using the following formula:

$Rrev=2^n-R-1$.

By modifying the Figure7 the initial state of an elementary CA is now composed of two successive configurations in the encryption scheme. The plain text is encoded as second state means $q_0$, while the first state is filled with the any random data or initial data which is generated by the Random Number Generator (RNG). The encryption is done by the forward iteration of the elementary CA. The final result is composed of two configuration and both of them must be used for both encryption and decryption. The first one is cipher text and second one is called final data. In the Decryption process the same operations are performed in reverse order. Encryption and Decryption Algorithms are shown in Figure 8 & Figure 9. The modified model is better than the previous model with respect to the circuit complexity, speed and cost.

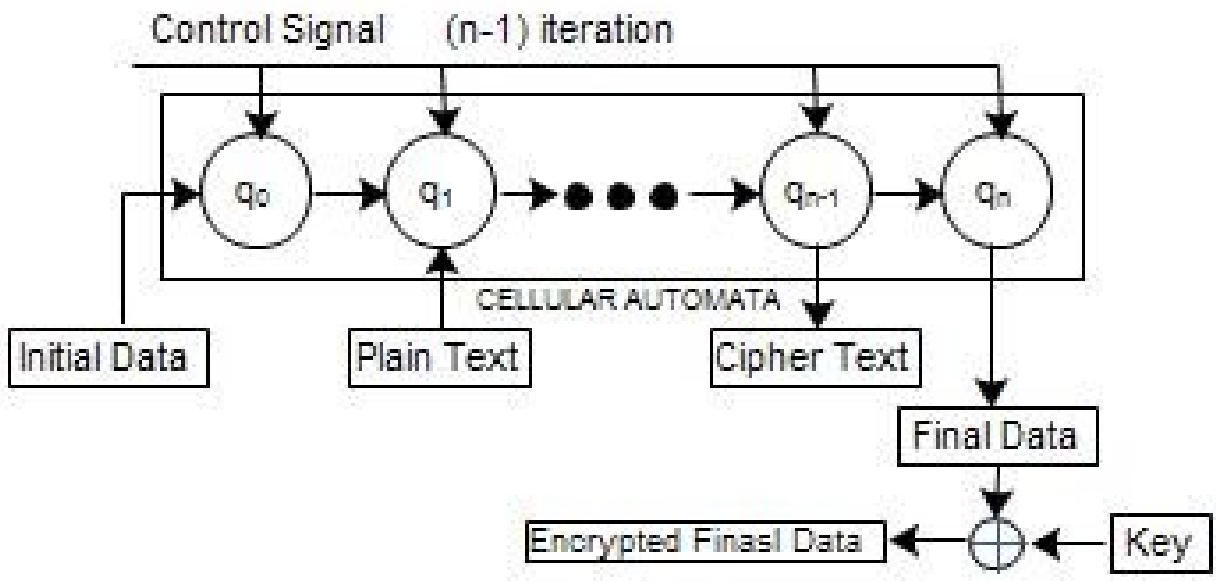

**Figure 8 Single Block Encryption Based on RPCA**

First encryption and decryption algorithm of a single plaintext block is shown. When using reversible PCA described in the previous section, plaintext is encoded as part of initial state of a PCA ($q_1$). Configuration $q_0$ is set up with some random data. Both configurations form an initial state of PCA. Encryption is done by forward iteration of CA by fixed number of steps according to some reversible rule. This process is shown Figure. 9.

Configuration on $q_{(n-1)}$ is a cipher text. The rule used during encryption is a secret key of that transformation. There are two options on how to treat configuration $q_{(n)}$ (called final data) generated by the encryption process. The most secure one assumes that this information is kept secret, which means that configuration $q_{(n)}$ becomes a part of the key. The disadvantage of this option is that the key changes with each encryption. This is because now the key is a function of a rule, plaintext and some initial data (Rid). In the second option the final configuration q(n) is encrypted using Vernam encryption algorithm. This is done by applying logical bitwise operation XOR on the final configuration $q_n$ and selected bits of the key.

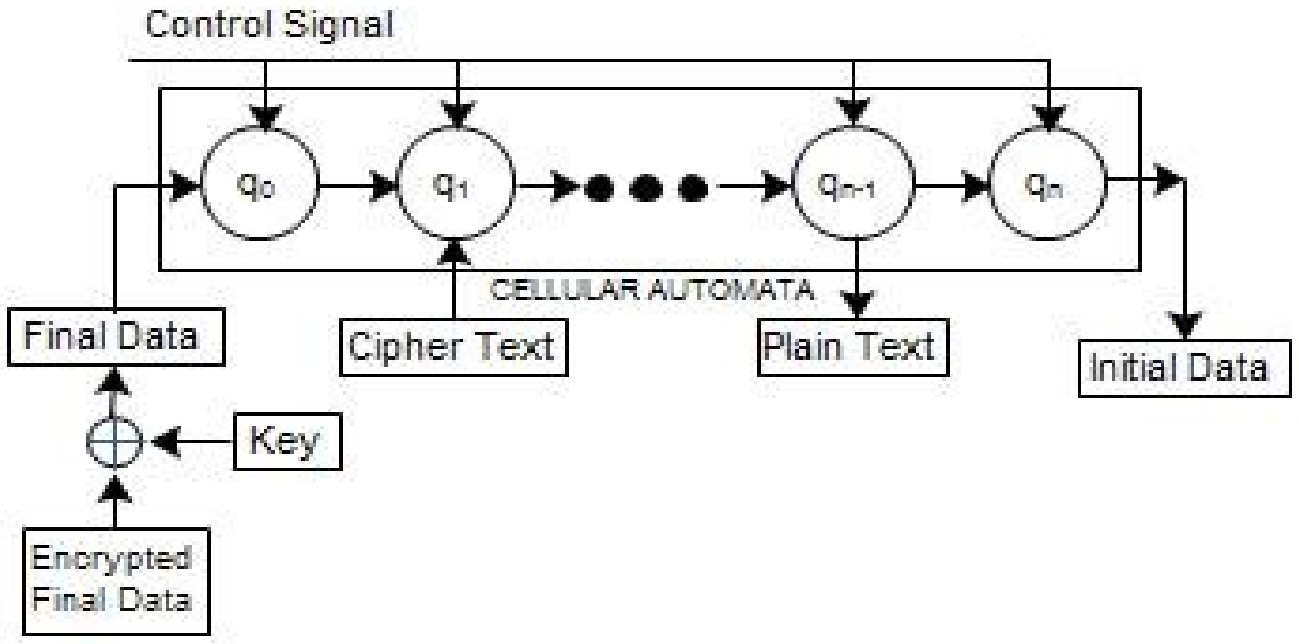

**Figure 9 Single block decryption using RPCA**

Decryption algorithm is shown on Figure. 8. The same operations as in encryption are used in reverse order. Initial state is composed of the final data and the cipher text. To obtain final data for the decryption, XOR operation must be applied first to encrypted final data and the key. Next, CA is iterated for the same number of steps as during encryption with use of the same secret rule.

The secret key has to include information about the rule number which is used for both encryption and decryption. The key would now be composed of a rule and final con figuration. By applying logical bitwise operation XOR operation on the final configuration and the key:

$ED_i=K_i$ XOR $F_i$ .
$ED_i$=ith Binary Digit of Encrypted final Data
$K_i$= ith Binary Digit of Key
$F_i$= ith Binary Digit of Final Data

### 4.2. Encryption for Reversible Programmable CA:

The rule combination for the Reversible programmable CA is based on Programmable CA. This Reversible PCA is used to provide real time keys for the block cipher.

Secret key is a rule for Encryption. $R_i$ is the Rule Encryption and $R_i$ is the Rule for Decryption.

### 4.2.1. Overview of Algorithm:

The algorithm is composed of three one dimensional CA labelled CAL, CAR and CAF. CAL and CAR are composed of 4 cells; CAF is composed of 128 cells. There are two input values for the encryption function: 128 bit plain text and 256 bit key. Automata CAL, CAR, and CAF are using the radius of 3 rules. All rules are belongs to the class of reversible rules. The encryption algorithm involves the four functions: 1) Byte Substitution 2) Row Shifting 3) Column Mixing 4) Add round column. Encryption is composed of n rounds which, means that all functions are applied n time on the plain text.

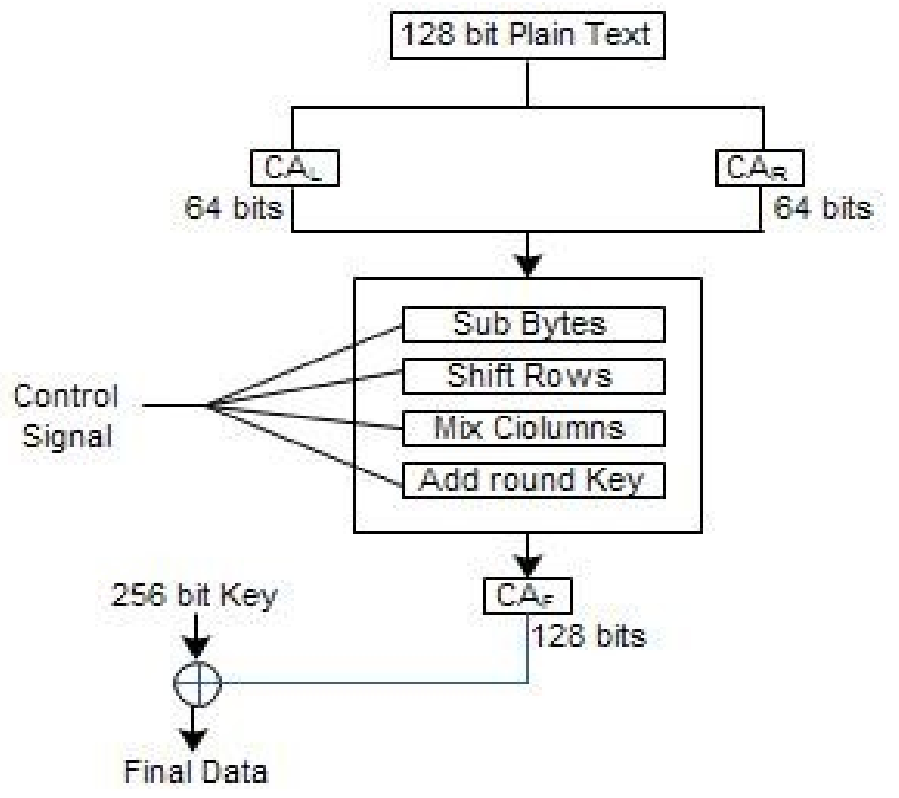

**Figure10 Proposed algorithm overview using RPCA**

### 4.2.2 Details of Single Round

The details of the single round of encryption are show on the Figure 6 each round starts with four 128bit values labelled q0, q1, q2 and q3. The left and right halves of each 64 bit values are divide into 32 bit parts labelled q0L, q0R, q1L q1R, q2L, q2R, q3L and q3R. The initial state of CAL is set up with values q0L, q1L, q2L, q3L and initial state of CAR with values q0R, q1R, q2R and q3R. The automata eight configuration are produced; q(n-3)L, q(n-2)L, q(n-1)L, qnL and q(n-3)R, q(n-2)R, q(n-1)R, qnR. The byte substitution operation configurations qnL and qnR, Row shifting operation configurations q(n-1)L and q(n-1)R, Column Mixing operation configurations q(n-2)L and q(n-2)R, add round key operation configurations q(n-3)L and q(n-3)R.
In Reverse rule of R is Rrev.

The reverse for the rule 15 is rule 240, rule 51 is rule 204 and rule 85 is rule 170. If two elementary CA rules are composing a reversible then one depend on each other. Knowing one of these rules then we can calculate the other one using the following formula:

$\mathbf{R_{rev}=2^n-R-1}$, where R is a given Rule, $\mathbf{R_{rev}}$ is a complement rule.

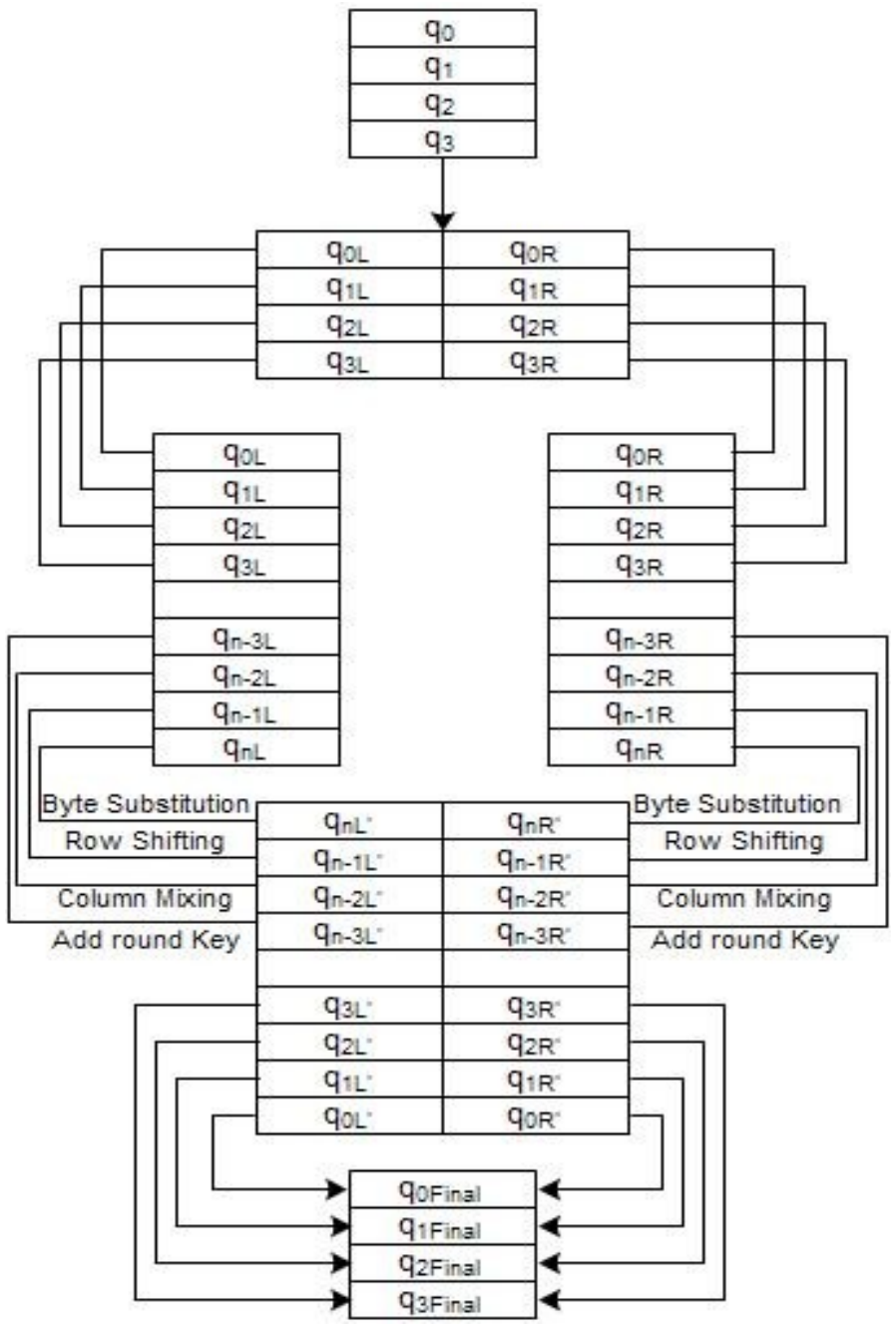

**Figure 11 Details of single round algorithm**

**Table 3: The rules that Updated The next state of the CA cells (reversible rule)**

| Rule | 111 | 110 | 101 | 100 | 011 | 010 | 001 | 000 |
|---|---|---|---|---|---|---|---|---|
| 15 | 0 | 0 | 0 | 0 | 1 | 1 | 1 | 1 |
| 240 | 1 | 1 | 1 | 1 | 0 | 0 | 0 | 0 |
| 51 | 0 | 0 | 1 | 1 | 0 | 0 | 1 | 1 |
| 204 | 1 | 1 | 0 | 0 | 1 | 1 | 0 | 0 |
| 85 | 0 | 1 | 0 | 1 | 0 | 1 | 0 | 1 |
| 170 | 1 | 0 | 1 | 0 | 1 | 0 | 1 | 0 |

The corresponding Combinational logic of rules 75 and 180 for elementary CA can be expressed as follows:

$$Rule15: a(i)(t+\ ) = 1 \oplus a(i-\ ))$$

$$Rule240: a(i)(t+\ ) = a(i-\ )(t))$$

$$Rule51: a(i)(t+\ ) = 1 \oplus\ (i)(t))$$

$$Rule204: a(i)(t+\ ) = a(i)(t))$$

$$Rule85: a(i)(t+\ ) = 1 \oplus\ (i)(t))$$

$$Rule170: a(i)(t+\ ) = a(i+\ )(t))$$

The encryption algorithm present in this paper is constructed using Reversible programmable CA based on rules 204, 240 and 170. The rules specify the evolution of the Reversible PCA from the neighbourhood configuration to the next state and these are represented in Table 3. Considering the RPCA with rule vector <204, 204, 240 and 170> under null boundary condition, and Corresponding truth table in simple in Table 4.

**Table 4: Selection of rules 204,240,170.**

| C1 | C2 | Rules Applied |
|---|---|---|
| 0 | 0 | 204 |
| 0 | 1 | 204 |
| 1 | 0 | 240 |
| 1 | 1 | 170 |

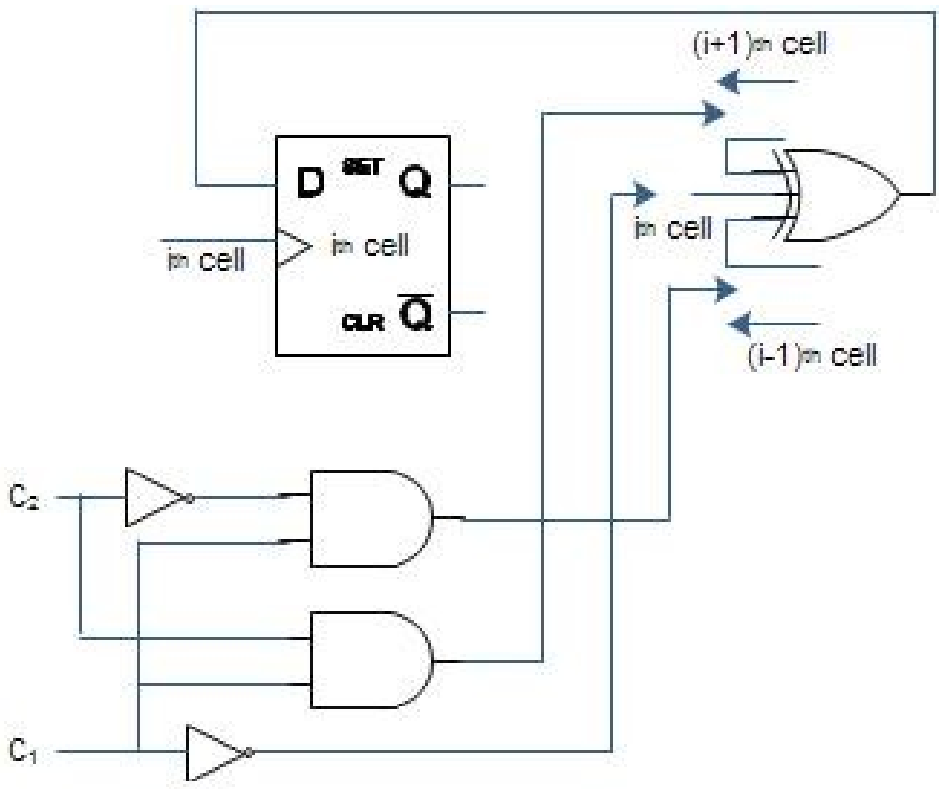

**Figure 12: RPCA structure with rule vector<204,204,240,170>**

A circuit with two control signals C1 and C2 are shown in Figure 12. When C1 is logic 0 then rule 204 is applied to the cell. When C1 is logic 1 and C2 is logic 0 then the cell is configure with rule 240. When C1 and C2 are both logic 1 then the cell configure with rule 170.

*Key:* Each CA is using its own reversible Rules. Each radius 3 rules is 64 and 128 bit length. This make key size is 256 bit. The Key structure is shown in Figure 13.

| | | |
|---|---|---|
| CAL Rule(64 bit) | CAR Rule(64 bit) | CAF Rule(128 bit) |

**Figure 13 Key Structure:**

*Decryption Process:* During the decryption process the same operation then the ones used for encryption is performed in reverse order.

*Algorithm Security:* There exist $2^{256}$ keys, which means a brute force attack appears practically impossible.

## 5 PERFORMANCE ANALYSIS

Block Encryption Algorithm Base on Reversible Programmable Cellular Automata can be implemented easily in both hardware and software. Each component of this cipher uses 3-neighborhood Programmable Cellular Automata, whose inherent parallelism feature makes parallel implementation nature .Simplicity and locality of the CA make it possible to build cheap and first devices containing large number of cells (processor ) working in parallel .The performance of the Block Encryption Algorithm based on Reversible Programmable Cellular Automata is compared with some of the Existing schemes and analyzed in this section.

### 5.1 Performance Analysis of Hardware Implementation

Since the Reversible Programmable Cellular Automata Based Components are regular, modular and cascade able structures the proposed scheme can be easily realize in hardware. To compare with existing scheme, the proposed scheme has been synthesized using Verilog targeted to Xilinx Vextex2. After synthesization it is found that ICEBERG [13] which needs 704 LTUs and Each round of AES Rijndael is even more critical as a single round their needs 2608 LUTs and its key round needs 768 LUTs [23]. The proposed model presented in this paper is implemented in hardware and found that the implementation cost is high because it uses control Programmable Cellular Automata.

### 5.2 Performance Analysis of Software Implementation:

The ICEBERG [13] scheme that proposed with the objective for efficient hardware implementation was not efficient for software implementation. The execution Speed of AES optimized code and the proposed non-optimized code on a Intel Core 2 Duo 2.0 GHZ, in open MP platform. The results are tabulated in Table 5.

**Table 5: Execution time for block encryption**

| Key Size | AES | Proposed Algorithm |
|---|---|---|
| 128 bit | 1.34 micro seconds | 1.01 micro seconds |
| 192 bit | 1.57 micro seconds | 1.21 micro seconds |
| 256 bit | 1.79 micro seconds | 1.43 micro seconds |

Implementation speed of our scheme was found to be faster than AES for all key sizes. This could be possible due to the inherited parallelism feature of CA.

## 6. CONCLUSION

This paper presents a new block encryption algorithm based on Reversible Programmable Cellular Automata theory. In Cellular Automata, rules are being selected to reduce the circuit complexity. This work ensures to generate $2^{256}$ potential keys which means that a brute force attack impossible. This algorithm also uses 128 bit block size which, implies an increase in security but may slow down the encryption/decryption process. This paper uses Reversible Programmable Cellular Automata because RPCA achieve high Parallelism, simple implementation and cheap cost.

## 7 REFERENCES

[1] S. Wolfram,"A new kind of science Wolfram Media", Inc.2002.

[2] Andrew Adamatzky, Ramon Alonso-Sanz, Anna Lawniczak, Genaro Juarez Martinez and Kenichi Morita,"AUTOMATA-2008 Theory and Application of Cellular Automata Thomas Worsch Editors", Inc-2008.

[3] Atual Kahata, "Cryptography and Network Security", 2nd Edition, Tata Mcgraw hill, 2008.

[4] W. Stallings ,"Cryptography and Network Security", 3rd edition, Prentice Hall,2003.

[5] Marcin Seredynski and Pascal Bouvary, "Block Cipher Based On Reversible Cellular Automata", New Generation Computing, 23(2005)245-258, Ohmsha Ltd and Springer.

[6] Anirban Kundu, Alok Ra jan Pal, Tanay sarkar, Moutan Banarjee, Sutirtha Kr. Guha and Deba jyoti Mukhopadhayay, "Comparative Study on Null Boundary and Periodic Boundary Neighbourhood Mulriple Attractor Cellular Automata for Classification", IEEE-2008.

[7] Guan P., "Cellular Automaton Public Key Cryptosystem", complex system 1, pp.51-56, 1987.

[8] S. Wolfram, "Cryptography with cellular Automata", pp.429-432, springer-1985.

[9] Somnath Tripathy and Sukumar Nandi, "LCASE: Lightweight Cellular Au- tomata Based on Symmetric Key Encryption, International journal of network security", Vol-8, No.2, PP.243-252, Mar-2009.

[10] P. Pal Chaudhuri, D. Roy Chowdhury, S. Nandi and S. Chattopadhyay, "Additive Cellular Automata Theory and Applications", Vol-1, IEEE Computer Society Press, USA, 1997.

[11] N Ganguly, "Cellular Automata Evolution: Theory and Applications in Pat- tern Classification", PhD thesis, B. E. College, 2003.

[12] Peter Anghelescu, Silviu Ionita, Emil Safron, "Block Encryption Using Hybrid Additive Cellular Automata", 7th International conference on Hybrid Intelligent Systems, IEEE-2007.

[13] F. Standaert ,G. Piret, G.Rouvroy , J. Quisquater, and J. Legat, "ICEBERG: An involutional Cipher efficient for block encryption in reconfigurable Hardware", LNCS 3017, pp. 279-299,Springer Verlag ,2004.

[14] S. Nandi,B.K.Kar and P. Pal Chaudhuri ,"Theory and Application of Cellular Automata in Cryptography",IEEE Transaction on computers,VOL.43.December 1994.

[15] Eduard Franti, Cristian Slav, T. Balan, M. Dascalu."Design of Cellular Automata Hardware for Cryptographic Application".IEEE-2004.

[16] N. Ganguly, B. Sarkar, A. Deutsch, G. Canright and P. chaudhuri, "a survey on Cellular Automata", Project BISON,IST-2001.

[17] S. Nandi, P. Chaudhuri, " Analysis of periodic and intermediate boundary 90/150 cellular Automata",IEEE-2005.

[18] Peter Anghelescu, Silviu Ionita, Emil Safron," FPGA Implementation of Hybrid Additive Programmable Cellular Automata", Eight International conference on Hybrid Intelligent Systems,IEEE-2008.

[19] Peter Anghelescu, "The projection and the analyses of the cellular automata for processing of information", Doctorial thesis (in Romanian),University of Pitesti,2007.

[20] Zhang Chuanwu, Peng Qicong, Li Yubo, "PCA Based Programmable Path signature Analysis In BIST", IEEE-2002.

[21] F. Maleki, A. Mohades, S. Mehdi hashemi, M.E. Shiri,"An Image Encryption System By Cellular Automata with Memory",The Third International conference on Availability,Reliablity and Security,IEEE-2008.

[22] M. Szaban, F. Seredynski, and Pascal Bouvry," Evolving Collective Behaviour of Cellular Automata for Cryptography",IEEE MELECON 2006.

[23] F. Standaert , G.Rouvroy , J. Quisquater, and J. Legat, " A methodology to implement block ciphers in reconfigurable hardware and its application to first and compact AES rijndael",FPGA-2003,pp-281-291, califonia, 2003.

[24] F. Seredynski, P. Bouvry, Albert Y. Zomaya, "Cellular Automata Computations and Secret Key Cryptography",Elsevier-2004.

[25] A. K. Das. ,"Additive Cellular Automata : Theory and Application as a Built-in Self-test Structure". PhD thesis, I.I.T.Kharagpur, India, 1990.

[26] A. K. Das and P. Pal Chaudhuri. "An Efficient On-Chip Deterministic Test Pattern Generation Scheme". Euromicro Journal, Microprocessing & Microprogramming, 26:195-204, 1989.

[27] A. K. Das and P. Pal Chaudhuri. "Efficient Characterization of Cellular Automata". Proc. IEE (Part E), 137(1):81-87,January 1990.

[28] A. K. Das and P. Pal Chaudhuri. "Vector Space Theoretic Analysis of Additive Cellular Automata and Its Applications for Pseudo-Exhaustive Test Pattern Generation". IEEE Trans. on Computers, 42(3):340-352, March 1993.

[29] A. K. Das, A. Sanyal, and P. Pal Chaudhuri. On Characterization of Cellular Automata with Matrix Algebra. Information Science, 61(3):251, 1991.

[30] S.Nandi, B.K. Kar,and P. Pal Chaudhuri, "Theory and Applications of Cellular Automata in Cryptography" , IEEE transactions on computers, vol. 43, no. 12, december ,1994.

[31] P. Anghelescu,S. Ionita and E. Sofron ,"Programmable Cellular Automata Based Encryption Algorithm",IEEE,2008.

[32] S. Das ," Theory and Applications of Nonlinear Cellular Automata In VLSI Design", PhD thesis, B. E. College, 2006.

[33] P. Maji and P. Pal Chaudhuri, "A Fuzzy Cellular Automata Based Pattern Classifier", DASFAA-2004, LNCS-2973, PP.494-505,2004.

[34] M. Seredynski and P. Bouvry ,"Block Cipher Based on Reversible Cellular Automata", New Generation Computer ,Springer-2005.

[35] E. R. Berlekamp, J. H. Conway, and R. K. Guy. "Winning ways for your Mathematical Plays", volume 2. Academic Press, 1984.

[36] Niloy Ganguly ,P. Maji, A. Das, B. K. Sikdar,&P. Pal Chaudhuri, "Characterization of Non-linear Cellular Automata Model for Pattern Recognition", AFSS 2002, LNAI 2275, pp. 214–220, Springer-2002.

[37] M. Bidlo and Z. Vasicek, "Gate Level Evolutionary Development Using Cellular Automata",IEEE-2008.

[38] Monica Dascglu, Eduard Franp and Zoltan Hascsi, "Hardware Version for Two-dimensional Cellular Automata",IEEE-1997.

[39] M. Mraz, N. Zimic, I. Lapanja, 1.Bajec," Fuzzy Cellular Automata: From Theory to Applications",IEEE-2000.

[40] F. B. Manning. "An Approach to Highly Integrated, Computer-maintained Cellular Arrays.", IEEE Trans. on Computers, C-26:536-552, 1977.

[41] P. Maji, C. Shaw, N. Ganguly, B. Sikdar, B. N. Roy, and P. Pal Chaudhuri. "Theory and Application of Cellular Automata for Pattern Classification. Fundamental Informaticae, ", 2003.

[42] P. D. Hortencius, R. D McLeod, W. Pries, D. M. Miller, and H. C. Card. "Cellular Automata based Pseudo-random Number Generators for Built-in Self-Test. IEEE Trans. on CAD", :842-859, August 1989.

**First A. Author** Mr Debasis Das received MCA degree from West Bengal University of Technology in 2008, currently pursuing M.Tech in Computer Science and Engineering from KIIT University, Bhubaneswar, Odisha, India. His area of interest includes Cellular Automata, Soft Computing, and Algorithm.

**Second B. Author** Mr Abhishek Ray received B.E. degree from Utkal University, Odisha in 1997, received M.E. degree from NIT, Rourkela, Odisha with specialization in Computer Science and Engineering in 2000. Worked as an Sr. Lecturer at Gandhi Institute of Engineering and Technology, Gunupur for a period of 5 years (March 2000 to June 2005). Currently working as Assistant Professor in School of Computer Engineering in KIIT University, Bhubaneswar, Odisha.